\documentclass[pra,preprint,amsmath,amssymb, superscriptaddress, longbibliography]{revtex4-1}

\usepackage{graphicx}
\usepackage{dcolumn}
\usepackage{bm}
\usepackage{multirow}
\usepackage{float}
\usepackage{wrapfig}
\setcitestyle{super}
\usepackage[normalem]{ulem}
\usepackage{color}
\usepackage{xcolor}
\usepackage[%
  colorlinks=true,
  urlcolor=blue,
  linkcolor=blue,
  citecolor=blue
]{hyperref}
\usepackage{xr}
\usepackage{booktabs}
\usepackage{longtable}

\usepackage{ulem}
\usepackage{verbatim}
\usepackage{color}

\begin{document}

\title{All-Optical Ultrafast Valley Switching in Two-Dimensional Materials}

\author{Navdeep Rana}
\affiliation{%
Department of Physics, Indian Institute of Technology Bombay,
           Powai, Mumbai 400076, India }
                      
\author{Gopal Dixit}
\email[]{gdixit@phy.iitb.ac.in}
\affiliation{%
Department of Physics, Indian Institute of Technology Bombay,
           Powai, Mumbai 400076, India }

\date{\today}



\begin{abstract}
Electrons in two-dimensional materials possess  an additional quantum attribute,
the valley pseudospin, labelled as  $\mathbf{K}$ and $\mathbf{K}^{\prime}$ -- analogous 
to the spin up and spin down. 
The majority of  research to achieve  valley-selective excitations in 
valleytronics 
depends  on  resonant circularly-polarised light with a given helicity. 
Not only acquiring  valley-selective electron excitation but also switching the 
excitation from one valley to another is quintessential  
for bringing  valleytronics-based technologies in reality. 
Present work introduces a coherent control protocol   
to initiate valley-selective excitation, de-excitation, and switch the excitation from one valley to another 
on the fly within tens of femtoseconds -- a  timescale faster than any valley decoherence time.
Our protocol is equally applicable to {\it both} gapped and gapless two-dimensional materials.   
Monolayer graphene and  molybdenum disulfide are used to test the universality. 
Moreover, the protocol is robust as it is insensitive to significant parameters of the protocol, such as 
dephasing times, wavelengths, and time delays  of the laser pulses. 
Present work goes beyond the existing paradigm of valleytronics, and opens a new realm of  
valley switch at PetaHertz rate. 
\end{abstract}

\maketitle

\section{Introduction} 
The successful synthesis of monolayer graphene has led to the proliferation in synthesizing of  
other two-dimensional (2D) materials, such as hexagonal boron nitride and transition-metal dichalcogenides~\cite{novoselov2005two, manzeli20172d, mak2010atomically}. 
These 2D materials exhibit attractive transport and optoelectronic properties, which 
hold promise for upcoming technologies~\cite{mak2016photonics}. 
One of the fascinating  features of these 2D materials is the electron's 
additional quantum attribute, the valley pseudospin -- analogous to the electron's spin. 
The valley pseudospin is associated with the valleys in the 
energy landscape of these materials.  
The hexagonal 2D materials are endowed with  two degenerate valleys 
situated at the corners of the Brillouin zone~\cite{xu2014spin, bussolotti2018roadmap}.
The flexible control over these valley pseudospins offers a platform 
to write, process, and store quantum information~\cite{mak2018light, ye2017optical}. 
Moreover, coherent switching of electron excitation from one valley to another,
on a timescale faster than the valley decoherence,   
is quintessential for valleytronics-based emerging quantum technologies at ambient conditions~\cite{vitale2018valleytronics, schaibley2016valleytronics}. 

\begin{figure}
	\includegraphics[width= \linewidth]{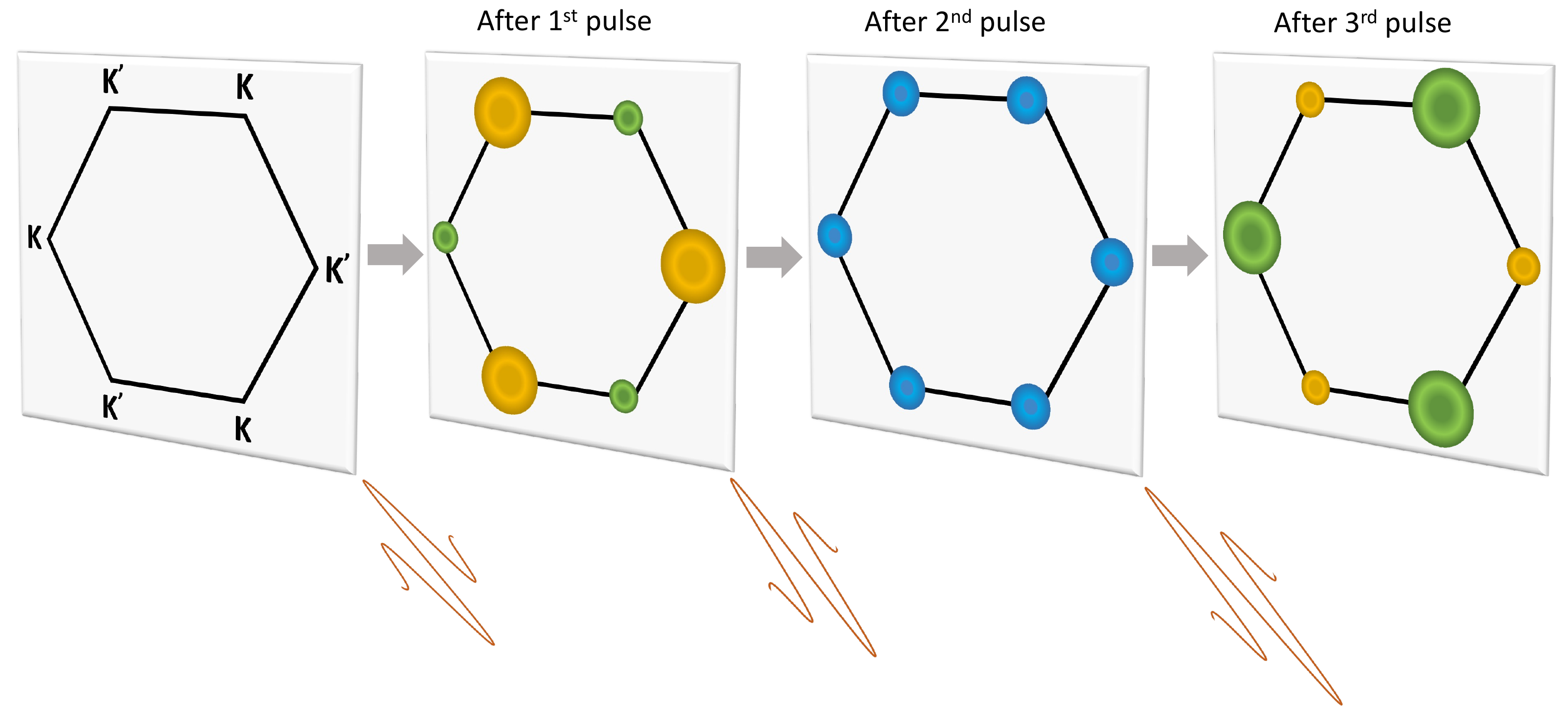}
	\caption{{Schematic of an all-optical ultrafast valley switch in a two-dimensional material.} 
		First panel: Brillouin zone of a 2D material along with high-symmetry points in momentum space. 
		Two degenerate valleys are located at the corners of the Brillouin zone, i.e., at $\mathbf{K}$ and $\mathbf{K}^{\prime}$. 
		Second panel:  valley excitation at  $\mathbf{K}^{\prime}$ (shown in yellow) is favored over 
		$\mathbf{K}$ (shown in green) after the end of the first  linear pulse. 
		Third panel:  the favored valley excitation  at  $\mathbf{K}^{\prime}$ is counterbalanced by 
		the second linear pulse subsequently  as  shown in blue.  
		Fourth panel:  reversal  in the preference of valley excitation at $\mathbf{K}$ (shown in green) over $\mathbf{K}^{\prime}$ (shown in yellow) by the action of the third linear pulse. 
		All three few-cycle pulses have well-defined carrier-envelope phase and 
		are  linearly polarised along $\Gamma - \mathbf{K}$ direction. } 
	\label{fig01}
\end{figure}

\begin{figure*}
	\includegraphics[width= \linewidth]{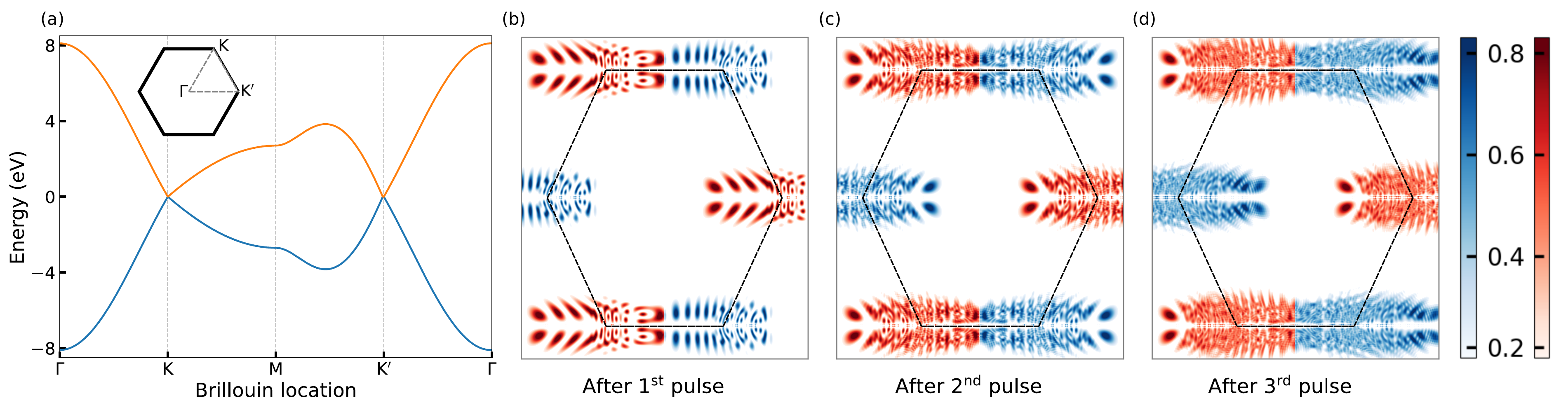}
	\caption{ All-optical valley switch in  graphene.
		(a) Energy band structure of graphene along high-symmetry
		points in the Brillouin zone. Two  inequivalent and degenerate valleys located at the corners
		of the Brillouin zone with crystal momenta $\mathbf{K}$ and  
		$\mathbf{K}^{\prime}$.  The first Brillouin zone is shown in the inset. 
		The momentum-resolved residual electronic  populations in the first Brillouin zone of the 
		conduction band are shown in 
		(b) - (d) after the end of the first, second, and third linear pulses, respectively. 
		(b) The first  pulse with CEP = 0$^{\circ}$ results  in asymmetric electronic  population, 
		leading  to preferred excitation around $\mathbf{K}^{\prime}$ valley. 
		(c)  After the second pulse with  
		CEP = 180$^{\circ}$, symmetric residual population in the Brillouin zone, i.e.,  
		$\mathbf{K}$ and  $\mathbf{K}^{\prime}$ valleys have equal population. 
		(d) Asymmetric  population distribution with more in the vicinity of  $\mathbf{K}$ valley. Thus, switch of  excitation from $\mathbf{K}^{\prime}$ to $\mathbf{K}$ valley after the end of the third pulse 
		with CEP = 180$^{\circ}$. Red and blue represent the residual populations around $\mathbf{K}^{\prime}$ and $\mathbf{K}$  valleys, respectively.}  
	\label{fig02}
\end{figure*}

Owing to the sign of the Berry curvature and resultant selection rules,  
left- or right-handed circularly polarized light, resonant to the material's band gap,  
has been employed to achieve  valley-selectivity excitation in gapped 2D materials~\cite{mak2012control, jones2013optical, gunlycke2011graphene, xiao2012coupled, zeng2012valley, cao2012valley, mak2014valley}.   
In recent years, nonresonant  laser pulses with tailored waveforms have been applied to obtain   
desired valley excitation in gapped 2D materials~\cite{motlagh2019topological, motlagh2018femtosecond, he2022dynamical, hashmi2022valley}.   
Similar significant success has been reported in gapless  2D materials
~\cite{gorbachev2014detecting,  yankowitz2012emergence,  grujic2014spin, settnes2016graphene, rycerz2007valley,  kundu2016floquet, kelardeh2022ultrashort}.
Huber and co-workers have experimentally achieved  
a milestone in coherent valley switching 
by using a combination of a pump pulse, resonant with the band gap in tungsten diselenide, 
and a controlled terahertz pulse~\cite{langer2018lightwave}. 
Recently, a scheme consisting of a pair of two time-separated 
orthogonally polarized linear pulses has been proposed
for valley excitation in gapped 2D materials~\cite{sharma2022valley}. 
Moreover, it has been shown that the desired control over valley excitation can be obtained by  
controlling the delay between the two pulses and their polarizations~\cite{sharma2022valley}. 
Ivanov and co-workers have proposed a control protocol to initiate and switch the valley excitation in gapped 2D materials using a sequence of four time-separated orthogonally polarized linear pulses~\cite{jimenez2022all}. 
Linear pulses, resonant with the band gap of the 2D materials, are used in 
both schemes~\cite{sharma2022valley, jimenez2022all}. 
Thus, they are unsuitable for gapless 2D materials, such as graphene. 
Moreover, both the schemes are susceptible to the relative polarizations and delay among the linear pulses. 
Thus, a robust and universal control protocol, applicable to gapped and gapless 2D materials, is lacking -- a major impediment in the realization of versatile light-based valleytronics devices. 

In this work, we provide a universal control protocol for initiating and switching the excitation from one valley
 to another on timescales faster than valley decoherence time.
To demonstrate universality, we apply our protocol to gapped and gapless 2D materials.  
Monolayer molybdenum disulfide (MoS$_2$) and graphene will be used to represent gapped and gapless 2D materials, respectively. 

The direction of electron flow is sensitive to the exact shape of the electric field waveform of the laser, 
which can be steered by the carrier-envelope phase (CEP) with attosecond precision. 
Recently, precise sculpting the waveforms of one and few-cycle laser pulses via  multioctave shaping has been demonstrated  experimentally~\cite{krogen2017generation, kawakami2020petahertz, savitsky2022single, higuchi2017light}. 
Linearly polarized laser pulses with CEP = 0$^{\circ}$ and = 180$^{\circ}$ have the electric field waveform 
opposite in direction.  
Thus, choosing few-cycle laser pulses with well-defined CEP  is expected to  facilitate switching
the excitation from one direction to another, which is the  essence of  our protocol. 

Let us understand how few-cycle CEP-stabilized linear pulse can induce valley polarization.  
It is known that the canonical momenta  is conserved at each instant of time during laser-solid interaction, i.e., ${\mathbf{k}(t_{2})} - {\mathbf{A}(t_{2})} =   {\mathbf{k}(t_{1})} - {\mathbf{A}(t_{1})}$ where ${t_{1}}$ and ${t_{2}}$ are two arbitrary instances during the laser pulse, $\mathbf{k}$ is the crystal momentum,
 and $\mathbf{A}(t)$ is the vector potential of the laser. 
The probability of an electron excitation  from valence to conduction band is maximum  at  
$\mathbf{K}$ and $\mathbf{K}^{\prime}$ valleys, i.e., near minimum band gap. 
Moreover, the likelihood of the excitation is maximum when the electric field strength of the laser is maximum.  
Thus, the maximum number of electrons are excited into the conduction band near the maxima of the electric 
field. Note that there is no preference for electron ejection between the two valleys as both valleys are 
energetically equivalent.

 As the laser pulse gets over, the electronic populations from the $\mathbf{K}$ and  $\mathbf{K}^{\prime}$ valleys are shifted to $\mathbf{k}_{1}$ and $\mathbf{k}_{2}$   as  $\mathbf{K} = \mathbf{k}_{1}(t \rightarrow \infty) + \mathbf{A}(t_{0})$ and $\mathbf{K}^{\prime} = \mathbf{k}_{2}(t \rightarrow \infty) + \mathbf{A}(t_{0})$ with $t_{0}$ as the time at which the electric field is maximum. 
The population distribution associated with the electron excitation is symmetric when the maxima of the electric field coincide with the zero of the vector potential, which is true for a longer laser pulse. 
Thus, ``relatively'' longer pulse yields symmetric  electron excitation around $\mathbf{K}$ and  
$\mathbf{K}^{\prime}$ valleys and, therefore, no valley polarization. 
This situation changes drastically as the pulse becomes shorter and sensitive to the CEP. 
In the case of the few-cycle CEP-sensitive pulse,  
the maxima of the electric field does not coincide with the zero of the vector potential. 
The value of the vector potential is sensitive to the value of the CEP. 
Thus,  the nonzero value of the vector potential at $t_{0}$ creates asymmetric population distribution around $\mathbf{K}$ and  
$\mathbf{K}^{\prime}$ valleys~\cite{mrudul2021controlling}.
The direction of the electron excitation is dictated by the direction of the field amplitude. 
For CEP value of 0$^{\circ}$, the magnitude of  the vector potential is positive at the maximum of electric field, which favors  $\mathbf{K}^{\prime}$  valley  over $\mathbf{K}$. 
On the other hand, as the CEP changes to 180$^{\circ}$, 
the magnitude of the vector potential becomes negative and thus $\mathbf{K}$  valley is favored. 

The key idea of our valley-switching protocol by an all-optical means in a 2D material, with two degenerate valleys at $\mathbf{K}$ and $\mathbf{K}^{\prime}$, 
is illustrated in  Fig.~\ref{fig01}. 
Coherent switching of an electron excitation from one valley to another in a 2D material can be 
envisioned as a three-step process: in the first step, 
 a linear laser pulse induces a preferable valley excitation at $\mathbf{K}^{\prime}$ over $\mathbf{K}$.  
An application of a second linear pulse during 
the second step could counterbalance the preferred valley excitation at $\mathbf{K}^{\prime}$ by 
reversing the direction of the excitation. 
The final step involves the action of a third linear pulse, similar to the one used in the second step, which would manage the excitation away from $\mathbf{K}^{\prime}$ valley as it favors the excitation to $\mathbf{K}$ valley.  
Thus, switching of the valley excitation from one to another can be potentially realized  by 
the action of three time-separated CEP-controlled linear pulses  polarized along the same direction. 
Our approach is simpler than the one discussed in Ref.~\cite{jimenez2022all} where a sequence of four orthogonally polarized linear pulses, resonant with the band gap of a 2D material, is used. 
Furthermore, as we  demonstrate below, our protocol is robust against several  limitations in the experimental setup as it is insensitive to the wavelengths of the laser pulses, time delay among the pulses, and dephasing time.  The protocol discussed in Ref.~\cite{jimenez2022all} is susceptible to the experimental setup.

\begin{figure*}
	\includegraphics[width=\linewidth]{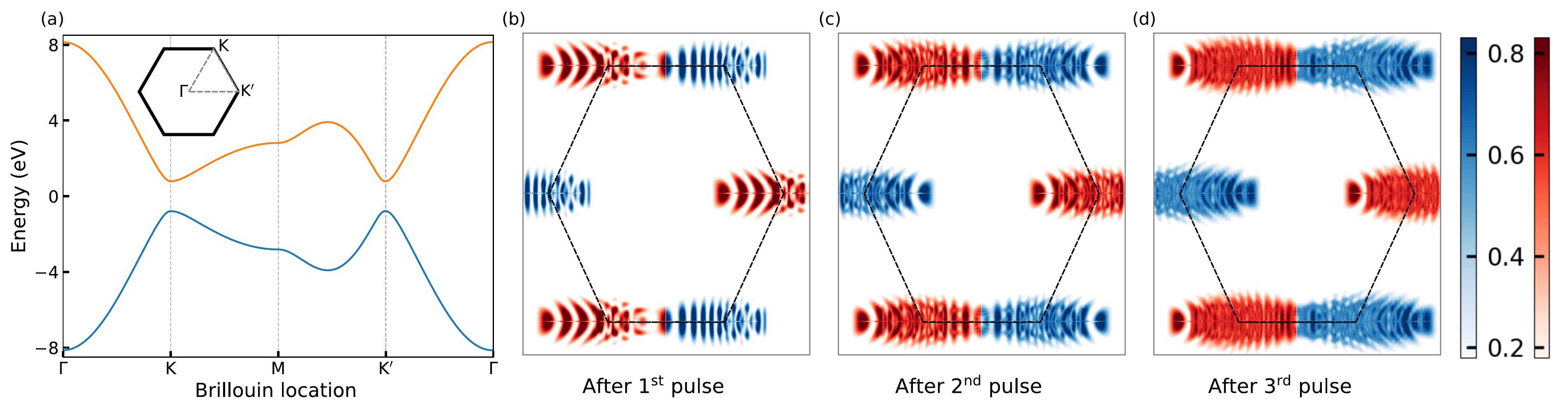}
	\caption{ All-optical valley switch in  MoS$_{2}$.  (a) Energy band structure of MoS$_{2}$ displays two valleys  along high-symmetry points at  $\mathbf{K}$ and $\mathbf{K}^{\prime}$. Energy band gaps 
		at $\mathbf{K}$ and $\mathbf{K}^{\prime}$ are 1.58 eV.  
		The momentum-resolved residual electronic  populations in the first Brillouin zone of the 
		conduction band after the end of the first, second, and third laser pulses are shown in (b) - (d), respectively.    
		Red and blue represent the residual populations around $\mathbf{K}^{\prime}$ and $\mathbf{K}$  valleys, respectively.
		The details of the linear pulses and the  residual populations in (b) - (d) are the same as in Fig.~\ref{fig02}.}
	\label{fig03}
\end{figure*}

\section{Computational Details}
The nearest-neighbour tight-binding approximation is used to obtain the electronic ground state of gapped and gapless graphene. The corresponding  Hamiltonian is expressed as
\begin{equation}
	\mathcal{{H}}_\textbf{k} =  -\gamma_{0}f^{*}(k)\hat{a}_\textbf{k}^{\dagger}\hat{b}_		           \textbf{k}-\dfrac{\Delta}{2}\hat{a}_\textbf{k}^{\dagger}\hat{a}_\textbf{k}+\dfrac{\Delta}{2}\hat{b}_\textbf{k}^{\dagger}\hat{b}_\textbf{k}+\text{H.c.}
     \label{eq:001}
\end{equation}
Here, $\gamma_0$ is the nearest-neighbor hopping energy, $\hat{a}_k^{\dagger} (\hat{b}_k)$  is creation (annihilation) operator corresponding to  the  A (B) type of the atom in the unit cell.
$f(k)$ is defined as $\sum_{i\in nn} ~e^{i\textbf{k}\cdot \textbf{d}_i}$ with 
$\textbf{d}_i$ as the separation vector between an atom with its nearest neighbor~\cite{mrudul2021high, mrudul2022high, rana2022high}. 
The band structure of the 2D material is obtained by solving Eq.~(\ref{eq:001}).  
To obtain the minimum band gap at $\mathbf{K}$  and $\mathbf{K}^{\prime}$ points, we use 
$\Delta = $ 0 and 1.58 eV for graphene and MoS$_{2}$, respectively.

Semiconductor-Bloch equations in Houston basis are used to simulate the electron dynamics  in the presence of few-cycle laser pulses~\cite{rana2022probing}  and defined as
\begin{subequations}
\begin{align}
    \frac{d}{dt}\rho_{vv}^{\textbf{k}} &= i\textbf{E}(t)\cdot \textbf{d}_{vc}(\textbf{k}_t)\rho_{cv}^{\textbf{k}} + \textrm{c.c.} \\
    \frac{d}{dt}\rho_{cv}^{\textbf{k}} &= -\left[i\varepsilon_{cv}(\textbf{k}_t)+\frac{1}{\textrm{T}_{2}}\right]\rho_{cv}^{\textbf{k}} + i\textbf{E}(t)\cdot\textbf{d}_{cv}(\textbf{k}_t)\left[\rho_{vv}^{\textbf{k}}-\rho_{cc}^{\textbf{k}}\right].
\end{align}
\label{SBE}
\end{subequations} 
Here, $\textbf{k}_t$ stands for $\textbf{k} + \textbf{A}(t)$,  $\textbf{E}(t)$ is 
the  electric field of the laser and $\textrm{T}_{2}$ represents a phenomenological dephasing term to incorporate  the decoherence between electrons and holes. 
 $\textbf{d}_{cv}(\textbf{k}) = i\langle c,\textbf{k} |\nabla_\textbf{k}|v,\textbf{k}\rangle$,  and  $\varepsilon_{cv}(\textbf{k})$ are the dipole matrix element  and band-gap energy between conduction and valence bands at $\textbf{k}$, respectively. 
 Note that $\rho_{cc}^{\textbf{k}}(t) = 1 - \rho_{vv}^{\textbf{k}}(t)$, and $\rho^{\textbf{k}}_{vc}(t) = \rho^{\textbf{k}*}_{cv}(t)$. The electronic population of the conduction band is estimated using $ n^{\textbf{k}}(t) =\rho^{\textbf{k}}_{cc}(t). $

To demonstrate the robustness of our idea, the parameters of the three laser pulses are similar 
for graphene and MoS$_2$. 
Realistic two-cycle pulses with well-defined CEP are used in our control protocol. 
All three pulses have a wavelength of 2.0 $\mu$m with 
cosine squared envelope, and are linearly polarised 
along $k_{x}$ direction, i.e., the zigzag ($\Gamma - \mathbf{K}$) direction. 
The CEP of the first, second, and third pulses are 0$^{\circ}$, 180$^{\circ}$, and 180$^{\circ}$, respectively. 
For graphene, the peak intensities of the first, second, and third pulses are
3.8$ \times 10^{12} $, 3.8$ \times 10^{12} $,  and 4.4$ \times 10^{12} $ W/cm$^2$, respectively. 
The peak intensities of the three pulses are slightly tuned to optimize valley switching in MoS$_2$. 
In this case, the peak intensities of the first, second, and third pulses are
2.6$ \times 10^{12} $, 2.6$ \times 10^{12} $, and 3.0$ \times 10^{12} $ W/cm$^2$, respectively. 
In both cases, the intensity of the third pulse is slightly higher than the second one to favor the 
electron  excitation to another valley. However, the ratios of the three pulses are kept constant  
in graphene and MoS$_2$. 
Note that the intensities of the laser pulses are  close to the damage threshold of monolayer graphene~\cite{roberts2011response}. By using an appropriate substrate,  
damage of the sample can be further avoided.   

\section{Results and Discussion}
Let us first discuss monolayer graphene's numerical results to confirm our valley switching protocol.  
There are two equivalent valleys at $\mathbf{K}$ and  
$\mathbf{K}^{\prime}$ as evident from the energy band structure of graphene along high-symmetry
points in the first Brillouin zone, see Fig.~\ref{fig02}(a). 
The band dispersion is linear, and energy gaps between valence and conduction bands at $\mathbf{K}$ and  
$\mathbf{K}^{\prime}$ valleys are zero. Thus, schemes based on resonant laser pulses for 
valley-selective excitation are not suited for graphene.    

The residual electronic populations in the conduction band after the first, second, and third pulses are presented in Figs.~\ref{fig02}(b) - \ref{fig02}(d), respectively. 
The populations in the vicinity of $\mathbf{K}$ and  
$\mathbf{K}^{\prime}$ are shown by red and blue, respectively. 
After the first pulse, the residual population is asymmetric. 
Moreover, the population in $\mathbf{K}^{\prime}$ valley is more in comparison to the $\mathbf{K}$ valley as reflected from Fig.~\ref{fig02}(b).

Let us uncover the underlying mechanism for the preferential population in $\mathbf{K}^{\prime}$ valley by the first laser pulse.  As discussed earlier,  the exact shape of the waveform of an intense laser dictates the direction of the electron's motion, either positive or negative, along the $x$ axis. 
Also, a sufficiently intense laser not only excites electrons in the valley but also its vicinity.
As the time starts, electrons are driven towards a positive direction as the vector potential of the first laser  has a negative slope and increases in the same direction. 
At some point in time, electrons change their direction of motion from positive to negative due to a change
in the slope of the vector potential  from negative to positive as the potential  peaks twice during each half cycle of the waveform.  
The two paths of the electron motion lead an interference, which results in 
interference fringes in the residual population as evident from Fig.~\ref{fig02}(b) [see 
S1 within the Supplemental  Material]. 
Due to zero CEP of the pulse,  the vector potential is asymmetric and has a net positive component, which leads the $\mathbf{K}^{\prime}$ valley to be more populated over $\mathbf{K}$ valley. 
Thus, the asymmetry of the laser waveform gets imprinted in the asymmetric residual populations. 

The preference for excitation from one valley to another can be altered by 
changing  the CEP from 0$^{\circ}$ to 180$^{\circ}$. 
To counterbalance the preferred population in $\mathbf{K}^{\prime}$ valley caused 
by the first pulse, 
let us employ a second pulse with CEP = 180$^{\circ}$ that favors the population in $\mathbf{K}$ valley. 
Thus, the combined effect of both pulses  leads to a symmetric residual population around both  
valleys as reflected from Fig.~\ref{fig02}(c). 

To break the equal population distribution around $\mathbf{K}$ and $\mathbf{K}^{\prime}$, 
a third pulse with CEP = 180$^{\circ}$ is applied, which 
induces asymmetric residual population. 
However, this time $\mathbf{K}$ valley has a more residual population over 
the $\mathbf{K}^{\prime}$ valley, see Fig.~\ref{fig02}(d). 
Thus, the action of three time-separated linear pulses switches the population from 
one valley to another completely~\cite{NoteX}. 
Moreover, the switch of valley excitation from $\mathbf{K}^{\prime}$ to  $\mathbf{K}$ valley occurs within a few tens of femtoseconds -- an order of magnitude 
faster than the valley decoherence time~\cite{vitale2018valleytronics, schaibley2016valleytronics}. 
Note that two-cycle laser pulses with CEP = 90$^{\circ}$ and 270$^{\circ}$ have zero net components and do not prefer any valley-selective excitation. 

After a successful illustration of valley switching in gapless graphene, let us explore the universality 
of our control protocol.
For this purpose, we  apply our protocol to MoS$_{2}$, which belongs to another class of 2D materials, i.e., gapped graphene. 
As evident from Fig.~\ref{fig03}(a), there are two valleys at $\mathbf{K}$ and $\mathbf{K}^{\prime}$ with a band gap of 1.58 eV, which makes MoS$_{2}$ a semiconductor in nature.  
To claim the universality, the sequence of the three time-separated laser pulses are the same for MoS$_{2}$ as for graphene. 

The residual population in the conduction band after the end of the first pulse is shown in
Fig.~\ref{fig03}(b). As evident from the figure, the first pulse with CEP = 0$^{\circ}$ favors 
electron excitation in $\mathbf{K}^{\prime}$ valley. 
The second pulse with CEP = 180$^{\circ}$ dispenses the population equally to both $\mathbf{K}$ 
and $\mathbf{K}^{\prime}$ valleys [see Fig.~\ref{fig03}(c)]. 
The valley excitation is completely switched to $\mathbf{K}$ valley by the action of the third  
pulse with CEP = 180$^{\circ}$ as reflected from Fig.~\ref{fig03}(d). 
The results presented in Fig.~\ref{fig03} unequivocally establish that our protocol with the same setup, as in graphene, is equally applicable to MoS$_{2}$ for a complete switch of valley excitation from one valley to another. 
Thus, our control protocol of valley switch is equally applicable to gapped and gapless 2D materials, unlike 
the protocol discussed in Ref.~\cite{jimenez2022all}.  

\begin{figure}{}
\includegraphics[width=0.7 \linewidth]{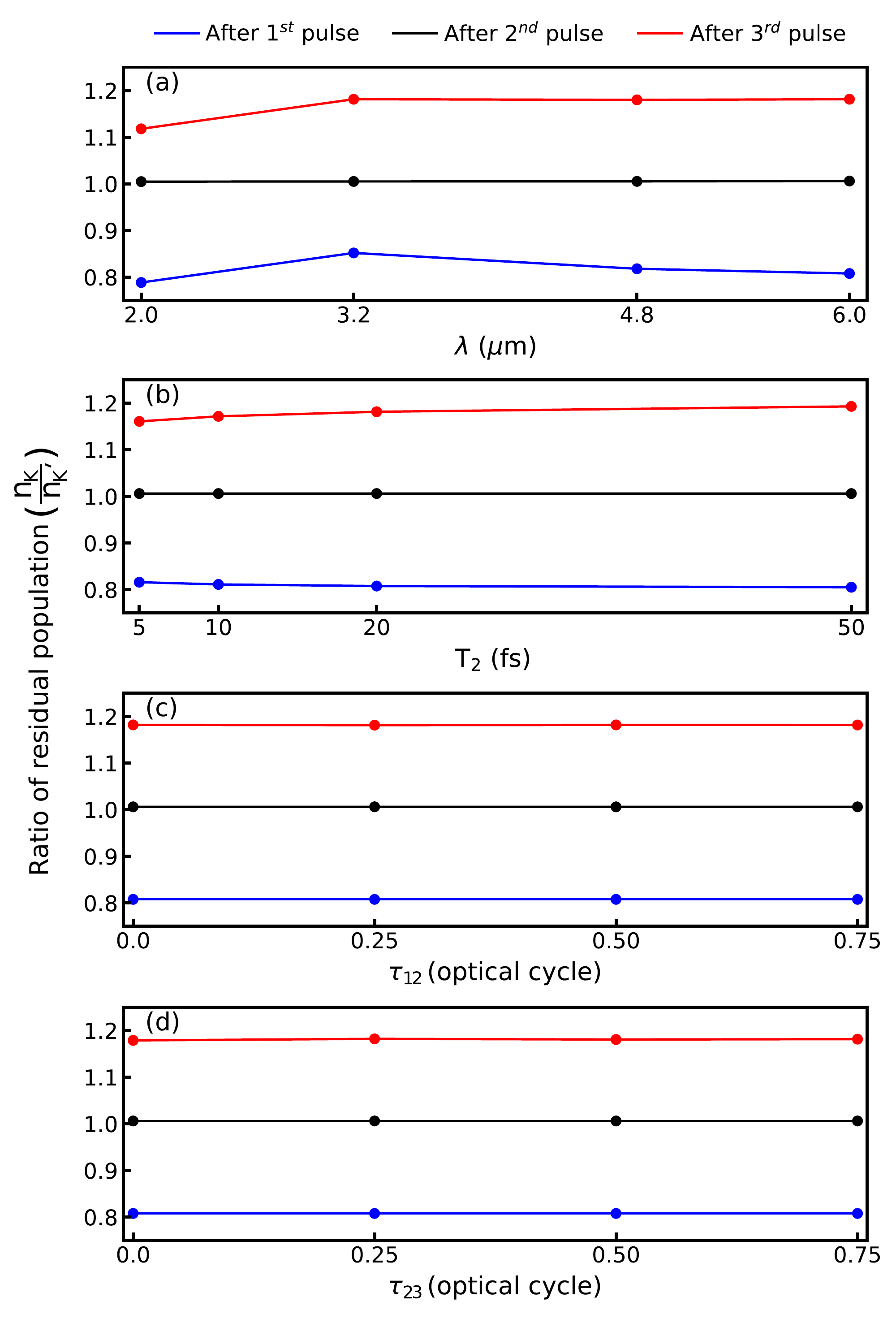}
\caption{Robustness of our protocol with respect to laser parameters and dephasing time.
The ratio of residual population in the conduction band 
around $\mathbf{K}$ 
and $\mathbf{K}^{\prime}$ valleys ($\textrm{n}_{\textrm{K}}/\textrm{n}_{\textrm{K}^{\prime}}$)  
with respect to (a) laser's wavelength $\lambda$, (b) dephasing time 
$\textrm{T}_{2}$, time delay between (c) the first and second laser pulses $\tau_{12}$, 
and (d) the second and third laser pulses $\tau_{23}$. 
Time delay between the second and third laser is constant in (c), and 
the delay between the first and second laser is constant in (d). 
The wavelength of all three pulses is kept the same in (a).
The residual populations after the first, second, and third pulses are shown by blue, black, and red, respectively. }
\label{fig04}
\end{figure}

After establishing valley switching in graphene and MoS$_{2}$, 
let us explore the robustness of our protocol concerning the various parameters in an experimental setup. 
To this end, we analyze the ratio of the residual population in graphene 
after the end of the first, second, and third laser pulses, as shown by blue, black, and red lines [See Fig.~\ref{fig04}]. 
As illustrated earlier, the residual population in the $\mathbf{K}^{\prime}$ valley is greater than the 
$\mathbf{K}$ valley, so the ratio is less than one. 
However, the ratio becomes one after the end of the second pulse as it counterbalances the population in both the valleys. 
The action of the third pulse favors the residual population in the $\mathbf{K}$ valley, and hence the ratio
becomes more than one. 
It is expected that the change in the wavelength of the laser pulses does not affect 
the outcome of the valley switch from $\mathbf{K}^{\prime}$ to $\mathbf{K}$ 
as the underlying mechanism relies on tuning the CEP of the pulses. 
The results in Fig.~\ref{fig04}(a) confirm our expectations as 
our protocol of the valley switch is insensitive with respect to the change in the wavelength
from 2.0  to 
6.0 $\mu$m for all the three pulses. 
Intensity ratios of the three pulses are kept constant while varying the wavelengths. 

It has been anticipated that 
the coherence between electron and hole at their respective energy bands is  a crucial factor 
in the functioning  of light-driven valleytronics devices~\cite{vitale2018valleytronics}. 
Thus, it is essential  to check how our  
protocol depends on the electron-hole coherence. 
The residual populations after the end of the three pulses are presented for different dephasing time 
$\textrm{T}_{2}$, which determines  the decoherence between electron and hole.
As evident from  Fig.~\ref{fig04}(b), the  ratios of the  populations remain 
unaffected for different values of  
$\textrm{T}_{2}$, ranging from 5 to 50 fs. 
Invariance of the ratios with $\textrm{T}_{2}$ also indicates that the electron excitation 
by CEP-controlled  laser pulses is dominated by the intraband motion~\cite{mrudul2021light}. 

It is natural to envisage that our control protocol is robust against the time delay among the three pulses as 
the populations ratio do not change with $\textrm{T}_{2}$. 
To verify this hypothesis, we present the population ratios for the different 
delays between the first and second pulses ($\tau_{12}$) while 
keeping the delay between the second and third pulses constant [see Fig.~\ref{fig04}(c)].
As evident from the figure, the residual populations after the end of the three pulses remain the same for different values of $\tau_{12}$.  
The same is true when the delay between the second and third pulses ($\tau_{23}$) 
is changed, while $\tau_{12}$ is kept constant [see Fig.~\ref{fig04}(d)].
Thus, the analysis of Figs.~\ref{fig04}(c) and \ref{fig04}(d) demonstrate   
the protocol of valley switching is independent of 
the time delay among the three pulses, i.e., changes in $\tau_{12}$ and $\tau_{23}$. 
The other parameters are kept  identical as in Fig.~\ref{fig02}, while varying  $\tau_{12}$ and $\tau_{23}$.

\section{Conclusion} 
In summary, we  illustrate  a coherent control protocol to switch the valley excitation from one valley to another in 2D materials on a timescale faster than any valley decoherence time. 
Our protocol consists of three time-separated two-cycle linear pulses, polarized along the same direction. 
For efficient valley switching, the electric field waveform of the employed pulses are tailored by tuning the CEP.  We  successfully illustrate the valley switch in graphene and MoS$_{2}$. 
Thus, our protocol is versatile as it relies on CEP-controlled nonresonant pulses and 
equally applicable to hexagonal gapped and gapless 2D materials.
Moreover, our protocol is resilient to significant parameters in an experimental setup as it is oblivious to the wavelength of the pulses, time delay among pulses, and the dephasing time. 
The laser pulses used in our protocol have been recently employed to explore 
electron dynamics in solids~\cite{kawakami2020petahertz, savitsky2022single, hui2022attosecond}. 
Therefore, our protocol of  valley switching is within reach of experimental feasibility.  
Additionally, the present work can be extended to realize logical operations using valley pseudospins -- similar to the recent experimental work in graphene using real and virtual charge carriers~\cite{boolakee2022light}. 
We  also test our protocol with CEP-controlled single-cycle  pulses, and our findings  
remain qualitatively the same.  However, valley polarization reduces drastically as the number of cycles in a laser pulse increases~\cite{mrudul2021controlling}.
To end, high-harmonic spectroscopy and time-resolved angle-resolved photoemission spectroscopy could be employed to read the outcomes of the valley switching~\cite{mrudul2020high, bharti2022high, ghimire2019,  reimann2018subcycle}.

\section{Acknowledgement}
We acknowledge fruitful discussions with  Misha Ivanov (MBI Berlin),  
Alvaro  Jimenez-Galan (MBI Berlin), Mandar Deshmukh (TIFR India), M. S. Mrudul (Uppsala University, Sweden) and  Sumiran Pujari (IIT Bombay).  
G. D. acknowledges support from SERB India 
(Project No. MTR/2021/000138).

\bibliography{solid_HHG}

\end{document}